\documentclass[showpacs,preprintnumbers,amsmath,amssymb,floatfix,prc,twocolumn]{revtex4}
\usepackage{graphicx}% Include figure files
\usepackage{dcolumn}% Align table columns on decimal point
\usepackage{bm}% bold math
\usepackage{color}

\def\beq{\begin{equation}}
\def\eeq{\end{equation}}
\def\be{\begin{eqnarray}}
\def\ee{\end{eqnarray}}

\newcommand{\lsim}{%less than or approx. symbol
 \mathrel{\setbox0=\hbox{$<$}\raise0.6ex\copy0\kern-\wd0
 \lower0.65ex\hbox{$\sim$}}}

\newcommand{\gsim}{%greater than or approx. symbol
 \mathrel{\setbox0=\hbox{$>$}\raise0.6ex\copy0\kern-\wd0
 \lower0.65ex\hbox{$\sim$}}}

\newcommand{\centsc}[1] {\multicolumn{1}{c}{#1}}%center single cell
\begin{document}

\title{Coulomb distortion and medium corrections in nucleon-removal  reactions}
\author{M. Karakoc$^{1,4)}$}
\author{A. Banu$^{2)}$}
\author{C.A. Bertulani$^{1)}$}
\author{L. Trache$^{3)}$}
\affiliation{$^{1)}$Department of Physics \& Astronomy,Texas A\&M University-Commerce, Commerce, TX 75428}
\affiliation{$^{2)}$Department of Physics \& Astronomy,James Madison University,Harrisonburg, VA 22807}
\affiliation{$^{3)}$Cyclotron Institute, Texas A\&M University, College Station, Texas 77843, USA}
\affiliation{$^{4)}$Department of Physics, Akdeniz University, TR-07058 Antalya, Turkey}
%\begin{abstract}
%We study  the effect of final state interactions and of medium modifications of the
%nucleon-nucleon (NN) cross sections on the nucleon knockout reactions. We compare the results obtained with  
%and without these effects to check their relevance in the extraction of spectroscopic factors.  
%Our results are compared to published experimental data for total nucleon-removal cross sections and 
%for momentum distribution of fragments. It is shown that final state interactions (mainly Coulomb distortion) 
%and medium effects leads to some relevant modifications of quantities extracted. 
%\end{abstract}

%\begin{verbatim}
\begin{abstract}
\begin{description}
\item[Background] One-nucleon removal reactions at or above the Fermi energy are important tools to 
explore the single-particle structure of exotic nuclei. Experimental data must be compared with 
calculations to extract structure information, evaluate correlation effects in nuclei or determine 
reaction rates for nuclear astrophysics. However, there is insufficient knowledge to calculate 
accurately the cross sections for these reactions.  
\item[Purpose] We evaluate the contributions of the final state interaction (FSI) and of the 
medium modifications of the nucleon-nucleon interactions and obtain the shapes and magnitudes of  
momentum distributions. Such effects have been often neglected in the literature. 
\item[Method] Calculations for reactions at energies 35 - 1000~MeV/nucleon are reported and compared 
to published data. For consistency, the state-of-the-art eikonal method for stripping and diffraction dissociation  is used.
\item[Results] We find that the two effects are important and their relative contributions vary with 
the energy and with the atomic and mass number of the projectile involved.
\item[Conclusions] These two often neglected effects modify considerably the one-nucleon removal 
cross sections. As expected,  the effect are largest at lower energies, around 50 MeV/nucleon and on heavy targets.
\end{description}
\end{abstract}
%\end{verbatim}


\maketitle

\section{Introduction}

Nucleon knockout reactions in nuclear collisions at and above the Fermi energy in nuclei have become 
an important tool  to determine the occupancy of single-particle states and the correlation effects 
in the nuclear many-body system (see, e.g. Refs. \cite{BM92,Gregers,Tostevin1999,han03,Gade2008a}). 
In peripheral, sudden collisions of fast-moving projectiles with a target nucleus, a single nucleon 
is removed from the projectile, producing projectile-like residues  in the exit channel, which are measured. 
Referred to the center-of-mass system of the projectile, the transferred momentum is ${\bf k}_c$. 
For the knockout reactions in the sudden approximation, this must equal the momentum of the struck nucleon
before the collision. The standard reaction models assume that the ground state of the projectile of 
spin and parity $J^{\pi}$ can be approximated by a superposition of configurations of the form 
$[I_c^{\pi_c}\otimes nlj]^{J^{\pi}}$ where $I_c^{\pi_c}$ denote the core states and the $nlj$ are the 
quantum numbers for the single particle wave functions in a spherical mean field potential. The 
measured partial cross sections to individual final levels $c$ of the core allow to extract,  by 
comparison with theoretical calculations, the spectroscopic factors for the individual core-single 
particle configurations. In complete analogy to the use of angular distributions in transfer 
reactions, the orbital angular momentum $l$ is revealed by the shape of the momentum distributions 
$P({\bf k}_c)$. It is obvious then that the accuracy of the extracted spectroscopic factors, a 
measure of the occupancy of the single-particle orbitals in nuclei, and the conclusions that may 
follow from these spectroscopic factors about correlations inside nuclei depend in direct measure on 
the accuracy of the cross section calculations. Similarly, the cross sections for one-proton removal 
reactions $X \rightarrow Y+p$ are directly related to the non-resonant part of the astrophysical 
S-factors for the inverse radiative proton capture Y(p,$\gamma$)X (see, e.g., 
Refs. \cite{trache01,trache02,han03,navratil06,BAN11}). Again, the results and their reliability 
depend directly on the reliability and accuracy of the reaction calculations 
(as discussed in \cite{trache04}).

The one-nucleon removal cross section is calculated in most reaction models as an incoherent sum of 
the contributions of all core-single particle configurations making the ground state of the fast moving projectile: 
\begin{equation}
\sigma_{-1n}=\sum S(c;nlj)\sigma_{sp}(nlj),
\label{eq1}
\end{equation}
where $S(c;nlj)$ and $\sigma_{sp}$ are the spectroscopic factors of each configuration and the single particle
removal cross section, respectively \cite{han03}. A similar relation is valid for momentum distributions. 
Systematic studies of projectiles and reactions allow the determination of the ordering, spacing and the
occupancy of orbitals, essential in assessing how nuclei evolve in the presence of large neutron or 
proton excess. Much was done in this respect in the last decade in various laboratories. 
This information can be compared to many-body nuclear structure calculations which are now able to reproduce
the measured masses, charge radii and low-lying excited states of a large number of nuclei. It was 
found that, e.g., for very exotic nuclei, the small additional stability that comes with the
filling of a particular orbital can have profound effects upon their existence as bound systems, their lifetime and
structure, and lead to the discovery of magic numbers that do not manifest along the valley of stability.

Extensions of the nucleon knockout formalism including the treatment of final-state interactions 
have been discussed in Ref. \cite{BH04} where it is shown that Coulomb final-state interactions are of
relevance. In the meantime, inclusion of higher-order effects \cite{AS00,AOS03} and a theory for
two-nucleon knockout \cite{Tostevin2006,Sim09,Simpson2009} have been developed. Knockout reactions 
represent a particular case for which higher projectile energies allow a simpler theoretical treatment of the reaction
mechanism, due to the simplicity of the reaction mechanism and the assumption of a single-step process.      

A microscopic approach to direct reactions uses an effective nucleon-nucleon (NN) interaction (e.g. 
those of Ref. \cite{Ra79}) to start with. This interaction is often used to construct  an optical 
potential  with its imaginary part  assumed to relate to the real part and its strength adjusted to 
reproduce experimental data. The real and imaginary parts of the potential can also be independent as 
in Refs. \cite{trache01,trache02}, where the procedure starts from a NN effective interaction with 
independent real and imaginary parts. For collisions at high energies ($E\gtrsim 100$), it is possible 
to show that instead of nucleon-nucleon interactions one can use nucleon-nucleon cross sections as 
the microscopic input \cite{HUS91}. In this case, an effective treatment of Pauli-blocking of 
nucleon-nucleon scattering is needed, as it manifests through medium modification of nucleon-nucleon 
cross sections. It is well known that medium modification of the nucleon-nucleon cross sections is 
necessary for an adequate numerical modeling of heavy-ion collision dynamics in central collisions. 
In these collisions, the ultimate purpose is to extract information about the nuclear equation of state (EOS) by
studying global collective variables describing the collision process. In direct reactions, such as 
one-nucleon removal reactions, medium effects of NN scattering are smaller because mostly  
low nuclear densities are probed.  A first study of this effect in knockout reactions was carried 
out in Ref. \cite{BC10}. Nonetheless, no comparison with experimental data was provided. In this work 
we explore further consequences of medium corrections and final state interactions in knockout reactions. 
We study  medium effects in the NN cross section in knockout reactions using the methods reported in 
Ref. \cite{BC10}, namely with a geometrical treatment of Pauli-blocking and  with the Dirac-Brueckner 
theory  in terms of baryon densities. We also explore the effect of final state interaction, in 
particular the effects of Coulomb distortion in the entrance and final reaction channels. This is of 
relevance as an increasing number of experiments use heavy targets with a large nuclear charge.
We compare our results of knockout cross section and momentum distribution calculations to a large 
number of published experimental data. The purpose is to improve the accuracy of the extracted 
spectroscopic factors that will lead to better understanding nuclear structure  and to check and 
improve the reliability of the use of one-nucleon removal reactions as indirect methods in nuclear 
astrophysics.

\section{Medium and distortion effects}
The geometrical treatment of Pauli corrections is performed using the isotropic NN scattering approximation 
because the numerical calculations can be largely simplified if we assume that the free nucleon-nucleon 
cross section is isotropic. In this  case, a formula which fits the numerical integration of the geometrical 
model  reads \cite{BC10} 
\begin{eqnarray}
\sigma_{NN}(E,\rho_p,\rho_t) &=&\sigma_{NN}^{free}(E)\frac{1}{1+{1.892\left(\frac{2\rho_<}{\rho_0}\right)\left(\frac{|\rho_p-\rho_t|}{\tilde{\rho}\rho_0}\right)^{2.75}}}\nonumber \\
&\times& 
\left\{
\begin{array}
[c]{c}%
\displaystyle{1-\frac{37.02 \tilde{\rho}^{2/3}}{E}}, \ \ \   {\rm if} \ \ E>46.27 \tilde{\rho}^{2/3}\\ \, \\
\displaystyle{\frac{E}{231.38\tilde{\rho}^{2/3}}},\ \ \ \ \  {\rm if} \ \ E\le 46.27 \tilde{\rho}^{2/3}\end{array}
\right.
\label{VM1}
\end{eqnarray}
where $E$ is the laboratory energy in MeV, $\tilde{\rho}=(\rho_p+\rho_t)/\rho_0$, $\rho_<={\rm min} (\rho_p,\rho_t)$, $\rho_{i=p, t}$ 
is the local density of nucleus $i$, and $\rho_0=0.17$ fm$^{-3}$. The parameters and models for the $\rho_p$ and $\rho_t$ 
densities which have been used to describe the nuclei in this work are presented in Table~\ref{grdens}.

The Brueckner method goes beyond the simple geometrical treatment of Pauli blocking. Some of the Brueckner 
results that we used in this analysis have been reported in  Refs. \cite{LI93,LI94}, where a simple 
parametrization was given. It reads (the misprinted factor 0.0256 in Ref. \cite{LI94} has been corrected to 0.00256)
\begin{eqnarray}
\sigma_{np}  &  =& \left[ 31.5 +0.092\left| 20.2-E^{0.53}\right|^{2.9}\right] \nonumber \\
&\times&\frac{1+0.0034E^{1.51} \rho^2}{1+21.55\rho^{1.34}}, \nonumber\\
\sigma_{pp}  &  = &\left[ 23.5 +0.00256\left( 18.2-E^{0.5}\right)^{4.0}\right]\nonumber \\
&\times&\frac{1+0.1667E^{1.05} \rho^3}{1+9.704\rho^{1.2}} . \label{brueckner}
\end{eqnarray}

The limits of validity of this parametrization are clearly associated with the limits of validity of 
the Brueckner calculations, which are valid only below the pion-production threshold.
A modification of this parametrization was introduced in Ref. \cite{Xian98} and consists in combining 
the free nucleon-nucleon cross sections parametrized in Ref. \cite{Cha90} with the results of Brueckner 
theory reported in Refs.  \cite{LI93,LI94}. 
\setlength{\tabcolsep}{0.05cm}
\begin{table}[hbtp]
\begin{center}
\begin{tabular}{clllll} \hline
Nuclei    &\centsc{Model}&\centsc{$<r_{ch}^2>^{1/2}$}&\centsc{$<r_m^2>^{1/2}$}&\centsc{a}&\centsc{$\alpha$}\\
          &         &\centsc{(fm)}&\centsc{(fm)}&\centsc{(fm)}&\centsc{(fm)}\\
\hline
(Target)  & \\
$^{9}$Be  &HO$^a$    & 2.50(9)   & 2.367 & 1.77(6)  & 0.631 \\
$^{12}$C  &HO$^b$    & -         & 2.332 & 1.584    & -     \\
\multicolumn{6}{l}{(Projectile - Core)}\\
$^{10}$Be &HO$^a$    & 2.50(9)   & 2.372 & 1.77(6)  & 0.631 \\
$^{14}$N  &HO$^a$    & 2.540(20) & 2.410 & 1.729(6) & 1.291 \\
$^{16}$B  &LDM$^{c}$ & -         & -     & -        & -     \\
$^{22}$Mg &HFB$^{d}$ & -         & 2.92  & -        & -     \\
$^{23}$O  &LDM$^{c}$ & -         & -     & -        & -     \\
$^{32}$Mg &HFB$^{d}$ & -         & 3.187 & -        & -     \\
\hline
\end{tabular}
\caption{Ground state densities are from Refs.~\cite{DE87,Khoa00,AUD03,GOR07,ANG04} where $r_{ch}$ 
and $r_m$ are root mean square radii of charge and nuclear matter densities, respectively.
$^{(a)}$The nuclear matter densities are obtained using the Harmonic-oscillator~(HO) charge densities 
with parameters a and $\alpha$ from Ref.~\cite{DE87} and the method in Ref.~\cite{Satch79}.
$^{(b)}$The HO nuclear matter density is from Ref.~\cite{Khoa00}.
$^{(c)}$LDM is Liquid Drop Model~\cite{Myers70}.
$^{(d)}$Hartree-Fock-Bogoliubov (HFB) calculations are from Refs.~\cite{AUD03,GOR07,ANG04}.}
\label{grdens}
\end{center} 
\end{table} 
Current theoretical models for the calculation of momentum distributions and cross sections in high-energy 
nucleon-removal reactions follow a semiclassical probabilistic approach, described, e.g., in 
Refs. \cite{HM85,hen96}. The method relies on the use of ``survival amplitudes" (or S-matrices) in 
the eikonal approximation, 
\begin{equation}
S_i(b)=\exp[i\chi(b)]=\exp\left[-\frac{i}{\hbar v}\int_{-\infty}^\infty U_{iT}({\bf r})dz\right], \label{sib}
\end{equation} 
where $r=\sqrt{b^2+z^2}$, and $U_{iT}$ is the particle($i$)-target($T$) optical potential. In Ref. \cite{HUS91},  
a relation has been developed between the optical potential and the nucleon-nucleon scattering amplitude. 
Such a relation is often referred in the literature as the ``t-$\rho\rho$ approximation". The  t-$\rho\rho$ 
approximation is the basis of most calculations of elastic and inelastic scattering involving radioactive nuclei, 
as experimentally deduced optical potentials are not often available. 
In this approximation, the eikonal phase becomes
\begin{equation}
\chi(b)=\frac{1}{k_{NN}}\int_{0}^{\infty}dq\ q\
\rho_{p}\left(  q\right) \rho_{t}\left(  q\right)  f_{NN}\left(
q\right)  J_{0}\left(  qb\right)
\ ,\label{eikphase}%
\end{equation}
where $\rho_{p,t}\left(  q\right)  $ is the Fourier transform of the nuclear
densities of the projectile and target, and $f_{NN}\left(  q\right)  $ is the
high-energy nucleon-nucleon scattering amplitude at forward angles, which can
be parametrized as
\begin{equation}
f_{NN}\left(  q\right)  =\frac{k_{NN}}{4\pi}\sigma_{NN}\left(  i+\alpha
_{NN}\right)  \exp\left(  -\beta_{NN}q^{2}\right)  \ .\label{fnn}%
\end{equation}

There are many ways of introducing final state interactions in direct nuclear reactions, some of which are discussed in Refs. 
\cite{BH04,BG06}. Besides Coulomb repulsion, included by modifying the straight-line trajectories 
accordingly, we have also modified the integral in Eq. \eqref{sib} by using the optical potential including 
the Coulomb potential that modifies the S-matrices according to $S_i=S_i^N\cdot S_i^C$. The 
Coulomb phase in $S_i^C$ is calculated by assuming a uniform charge distribution with radius $R$, and is given by
\begin{eqnarray}
\chi_{C}(b)  &  =&2\eta\Big\{  \Theta(b-R)\ \ln(kb)\nonumber \\
&+&\Theta(R-b)\left[
\ln(kR)+\ln(1+\sqrt{1-b^{2}/R^{2}})\right.   \nonumber\\
&& \left. \left.    -\sqrt{1-b^{2}/R^{2}}-{\frac{1}{3}}(1-b^{2}/R^{2}%
)^{3/2}\right]  \right\}  \ ,\label{chico_0}%
\end{eqnarray}
where $\Theta$ is the step function. The value $R$ is chosen to be small enough so 
that the nuclear S-matrices are basically zero below $b=R$ because of the strong absorption at small impact parameters.

\section{Results and discussion}
In this section the results for momentum distributions and nucleon-removal cross sections 
are compared to several experimental data. The focus is the importance of medium 
corrections of nucleon-nucleon cross sections and of Coulomb distortions. Both effects 
are expected to decrease as the bombarding energy increases. 
It is important to include such effects in order to minimize the uncertainty in the 
extraction of spectroscopic factors, especially at low bombarding energies. To substantiate 
this assertion, we analyze low energy data on knockout reactions and compare to high energy data.
In order to identify the separate contribution of these two factors, we do not vary the geometry of 
the nucleon binding potentials used to calculate the single particle radial wave functions. That was 
identified in the literature as another major factor in the calculations, leading to large variations 
in the extracted spectroscopic factors.

\begin{figure}[t]
\includegraphics[scale=0.4,keepaspectratio=true,clip=true]{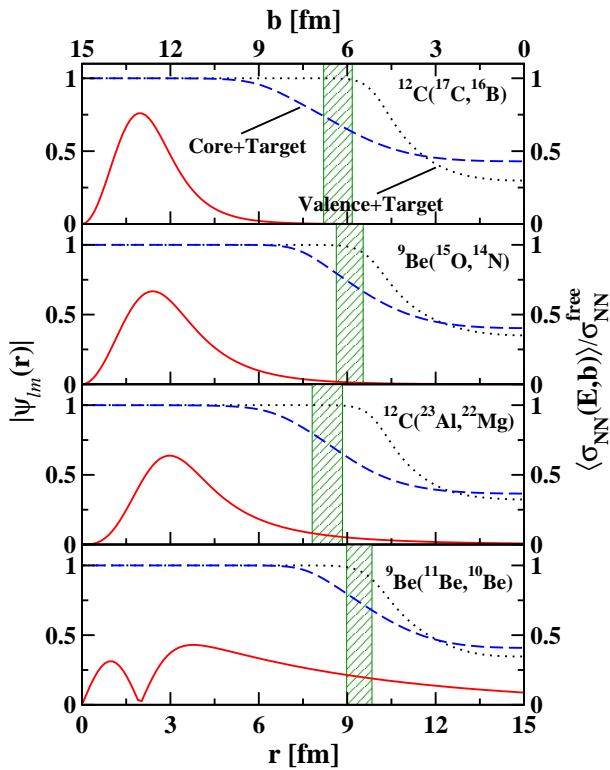}
\caption[Fig. 1]{(Color online). {\it Top and right scale:} Ratio between average in-medium  and 
the free nucleon-nucleon cross section as a function of the impact parameter. Dashed and dotted 
curves are for core-target and valence nucleon-target average nucleon-nucleon cross sections, 
respectively. Notice that the target center of mass is located on the right of the top axis scale.
The shaded areas represent the strong absorption radii where the knockout reactions most likely 
occur and $r_{sa}=b_{sa}=(1.1\pm0.1)~(A_P^{1/3}+A_T^{1/3})$~fm.
{\it Bottom and left scale:} Radial wave functions in arbitrary units (solid curves) for the valence 
nucleon-core system and for a few representative reactions considered in this work. We have taken 
only one configuration in cases of systems with multiple configurations.}
\label{Fig01}
\end{figure}

\begin{figure}[t]
\includegraphics[scale=0.35,keepaspectratio=true,clip=true]{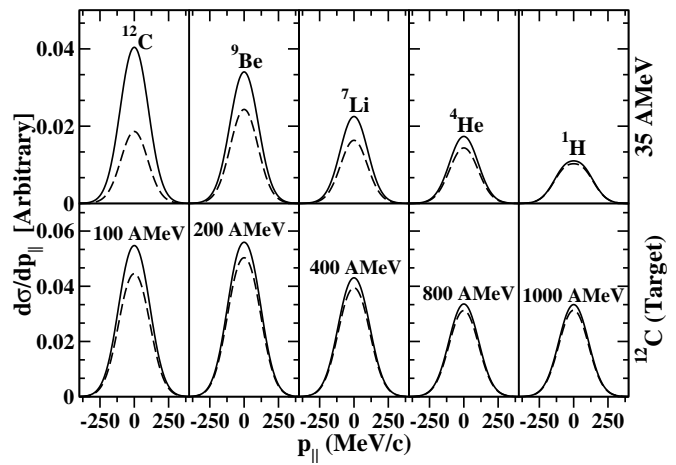}
\caption[Fig. 2]{{\it Upper panel}: Display of Coulomb scattering effects in longitudinal momentum 
distributions for the reaction T($^{17}$C,$^{16}$B)X at 35~MeV/u as a function of target T. 
The solid (dashed) curves are calculations with (without) Coulomb distortion. {\it Lower panel}: 
Same as above, but for  $^{12}$C target and for different beam energies. }
\label{Fig02}
\end{figure} 

The relevance of medium corrections are motivated by the effect summarized in Figure  \ref{Fig01}.
The cases are shown in Figure \ref{Fig01} are for $^{12}$C($^{17}$C,$^{16}$B) at 35~MeV/u, $^{9}$Be($^{15}$O,$^{14}$N) at 56~MeV/u, 
$^{12}$C($^{23}$Al,$^{22}$Mg) at 50~MeV/u and $^{9}$Be($^{11}$Be,$^{10}$Be) at 60~MeV/u.
Later we will discuss more results for each system separately.
The dashed and dotted curves show the ratio between the average in-medium  and the free nucleon-nucleon 
cross section as a function of impact parameter (see top scale). We define the average nucleon-nucleon 
cross section at the distance of closest approach between the projectile and the target using 
Eq. \eqref{VM1} and the definition
\begin{equation}
\left\langle \sigma_{NN} (E,b) \right\rangle = \frac{\int d^3 r_p \ \rho_p({\bf r_p}) \rho_t({\bf r_p}+{\bf b}) \ \sigma_{NN} (E, \rho_p,\rho_t)}{\int d^3 r \rho_p({\bf r_p}) \rho_t({\bf r_p}+{\bf b})} , \label{avesigEb}
\end{equation}
where ${\bf b}$ is the impact parameter vector, perpendicular to the beam axis. 
  
In Figure \ref{Fig01}, the dashed and dotted curves are for core-target and valence nucleon-target 
average nucleon-nucleon cross sections, respectively. Notice that the target center of mass is located 
on the right of the top axis scale. Also shown in the figure are the radial wave functions (solid curves) 
for the valence nucleon-core system and for a few representative reactions considered in this work. 
For simplicity, in this figure we have used only one of the main configurations for the projectile ground 
state (more detailed and complete calculations are reported later on in this section). 
The binding energies ``effectively" decrease from the top to low panels. We mention ``effectively" because, 
although the binding energy of the valence proton in $^{23}$Al is  smaller than for the valence neutron 
in $^{11}$Be, the Coulomb barrier creates an effectively larger binding in $^{23}$Al.  

It is clearly noticeable from Figure~\ref{Fig01} that the wavefunctions of weakly  bound systems 
extend far within the target where the nucleon-nucleon cross sections are strongly modified by 
the medium. We have to emphasize that the shaded areas in Fig.~\ref{Fig01} are relevant
to stress the importance of medium effects at surface region since the reaction is peripheral
due to strong absorption at $b<b_{sa}$. Momentum distributions and nucleon removal cross sections in knockout 
reactions are thus expected to change appreciably with the inclusion of medium corrections of 
nucleon-nucleon cross section. Such corrections are also expected to play a more significant role 
for loosely-bound systems.

In the following, we discuss Coulomb corrections. Here we consider the simplest and most straightforward 
correction one can do, namely the inclusion of a Coulomb phase, which accounts for the distortion of the elastic scattering 
of the core fragment.   It has been usually taken for granted that longitudinal momentum distributions are 
little affected by elastic scattering of the core fragment, the reason being that the longitudinal forces 
acting on the core fragment reverse sign as the projectile passes by the target, leading to a reduced 
distortion effect \cite{BM92}. Further, as has been  shown in Ref. \cite{BH04}, the transverse momentum 
distributions are strongly affected by both nuclear and Coulomb elastic scattering. For heavier targets 
the distortions are predominantly due to Coulomb repulsion \cite{BH04}.   It is worthwhile mentioning that the implications of the
findings on Coulomb distortion effects presented in Ref. \cite{BH04} have been neglected in the literature. In order to avoid dealing with 
the effects of the Coulomb scattering, experiments are usually performed with light targets, such as $^9$Be 
and relatively high energies, $E\gtrsim 50$~MeV/nucleon.  In this work we show that 
these argumentations are not always valid and need to be studied with care.

As discussed in the previous section, in the presence of the Coulomb field the eikonal S-matrices 
factorize as the product of the nuclear and the Coulomb contributions: $S(b)=S_n(b) S_C(b)$. Although 
this does not make any difference for the total stripping cross sections (see Eq. (20) of Ref. \cite{BC10}), 
it has an impact on the diffraction dissociation cross section (through the second term of Eq. (21) of 
Ref. \cite{BC10}). This means that not only transverse, but also  longitudinal momentum distributions 
will be affected by the Coulomb field. This is shown in Figure \ref{Fig02} for the longitudinal momentum 
distributions of several systems which we will consider in details later in this section.
It is evident from the upper panels of this figure that longitudinal momentum distributions in knockout 
reactions T($^{17}$C,$^{16}$B)X (and their total cross sections) are strongly influenced by the Coulomb 
field of the target T at bombarding energies of 35~MeV/nucleon. The solid (dashed) curves are calculations 
with (without) the inclusion of Coulomb scattering. It is also evident that even for the case of light 
targets, such as $^9$Be and $^7$Li, the distributions change appreciably. The lower panels show calculations 
for the same reaction, but for  $^{12}$C targets and as a function of the bombarding energy. It is clear 
that distortions are important even for usually considered ``safe" energies, such as 100~MeV/nucleon.
\begin{figure}[t]
\includegraphics[scale=0.35,keepaspectratio=true,clip=true]{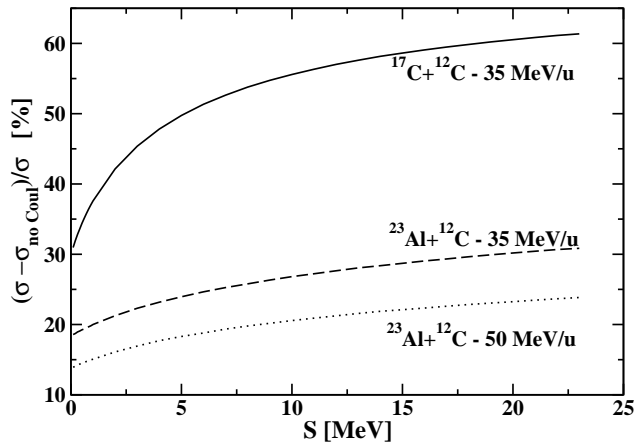}
\caption[Fig. 3]{Total nucleon removal cross sections for the reactions $^{12}$C($^{17}$C,$^{16}$B)  
(solid curve) and $^{12}$C($^{23}$Al,$^{22}$Mg) (dashed curve)
at 35~MeV/u. We artificially vary the separation energy $S$ of the proton in $^{17}$C and in $^{23}$Al. 
The dotted curve shows the  calculation for $^{12}$C($^{23}$Al,$^{22}$Mg) at 50~MeV/u.}
\label{Fig03}
\end{figure}

We found that the effect of Coulomb scattering is relatively larger for systems with smaller sizes. 
This is illustrated in Figure \ref{Fig03} where we present our calculations for the total nucleon 
removal cross sections for the reactions $^{12}$C($^{17}$C,$^{16}$B)  (solid curve) and 
$^{12}$C($^{23}$Al,$^{22}$Mg) (dashed curve) at 35 MeV/u. We artificially vary the binding energy 
of the proton in $^{17}$C and in $^{23}$Al. As the separation energy increases so does the percent 
difference of the cross section, $(\sigma-\sigma_{\rm no \ Coul})/\sigma$, where   $\sigma_{\rm no \ Coul}$ 
is the cross section without the Coulomb scattering phases. With increasing separation energy  
the relative valence nucleon-core distance decreases and  the nucleon removal cross section decreases, 
but the relative importance of the Coulomb scattering increases. For very small energies the effects 
of Coulomb dissociation (not considered here) should also become relevant and increase the magnitude 
of the cross sections. The relative importance of the Coulomb scattering for the removal cross sections 
decreases with the bombarding energy. This is shown in Figure \ref{Fig03} with the calculation for 
$^{12}$C($^{23}$Al,$^{22}$Mg) at 50 MeV/u (dotted curve). 

These preliminary discussions support our conclusion that both medium effects and Coulomb distortion play a 
relevant role in knockout reactions. Next we consider a  series of published data for which neither 
medium or Coulomb corrections were included. We thus quantify the changes in the extracted values of 
spectroscopic factors in case these effects were to be included in the experimental analysis.
 
\subsection{$^{12}$C($^{23}$Al,$^{22}$Mg)X at 50~MeV/u} 
Recently, the $^{12}$C($^{23}$Al,$^{22}$Mg)X knockout reaction has been studied at
50~MeV/nucleon to investigate the ground state properties of $^{23}$Al \cite{BAN11}.
It was shown that the ground-state structure of $^{23}$Al is a configuration 
mixing of a \textit{d}-orbital valence proton coupled to four core states of $^{22}$Mg - 
0$^{+}_{gs}$, 2$^{+}_{1}$, 4$^{+}_{1}$, 4$^{+}_{2}$. The ground state 
spin and parity of $^{23}$Al as $J^{\pi} = 5/2^{+}$ has been confirmed.
This experiment had the advantage that exclusive measurements were done and momentum distributions were 
determined for the four major configurations in the ground state of the projectile ($^{23}$Al).
\begin{figure}[ht]
\includegraphics[scale=0.40,keepaspectratio=true,clip=true]{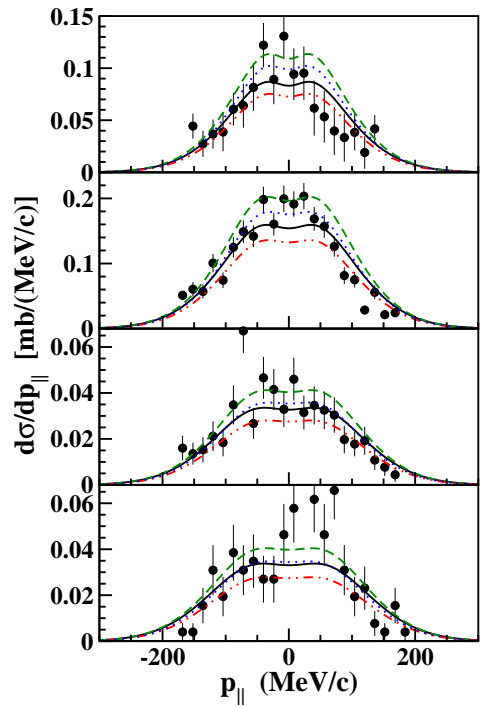}
\caption[Fig. 4]{(Color online). Comparison of experimental data of Ref. \cite{BAN11}
and calculations for exclusive longitudinal momentum distributions in the knockout reaction $^{12}$C($^{23}$Al,$^{22}$Mg)X  
at 50~MeV/nucleon. The solid line has both Coulomb and medium corrections. The dashed-curve has no 
medium corrections. The dashed-dotted line includes calculations without Coulomb corrections. The 
dotted curve neither includes medium effects nor Coulomb corrections.}
\label{Fig04}
\end{figure}

In this work, we have analyzed the $^{12}$C($^{23}$Al,$^{22}$Mg)X system to check the relevance of Coulomb and
medium effects. The $1d_{5/2}$ wave functions for the valence proton were generated in a spherical 
Woods-Saxon (WS) potential with the parameters given in Table \ref{bound}. 

In the optical limit of the Glauber theory and the 
t-$\rho\rho$ approximation (explained in detail in Refs.
\cite{Ra79,HUS91}), the eikonal phase is obtained from the input of the nuclear ground state
densities and the energy dependent nucleon-nucleon cross sections. The ground state density parameters 
and models used in this work are shown in Table \ref{grdens}, and our results are 
presented in Fig.~\ref{Fig04} and Table  \ref{t_preAl23}.
\begin{table*}[!]
\begin{center}
\setlength{\tabcolsep}{0.25cm}
\begin{tabular}{ccccccc} 
\hline
                                                 & $ E_x$&\multicolumn{4}{c}{$\sigma_{sp}(nlj)$ [mb]} \\
                 &                &\multicolumn{4}{c}{$\overbrace{\rule{6cm}{0pt}}$}\\[-5pt]
    Configuration& [keV]          & Full   &  no medium  &  no Coulomb  & Free &    \\
\hline
$^{22}{\rm Mg}(0^{+}_{gs})\otimes \pi_{1d_{5/2}}$&            0    &    27.1  & 33.8  & 23.2  & 30.1  \\
$^{22}{\rm Mg}(2^{+}_{1})\otimes \pi_{1d_{5/2}}$ &         1247    &    23.7  & 28.7  & 19.9  & 25.1  \\
$^{22}{\rm Mg}(4^{+}_{1})\otimes \pi_{1d_{5/2}}$ &         3308    &    20.4  & 23.9  & 16.7  & 20.4  \\
$^{22}{\rm Mg}(4^{+}_{2})\otimes \pi_{1d_{5/2}}$ &         5293    &    18.4  & 21.0  & 14.8  & 17.6  \\
\hline
\end{tabular}
\caption{Single particle cross sections are shown for each case separately.}
\label{t_preAl23}
\end{center}
\end{table*} 

To understand the effects of medium and Coulomb corrections, we have performed 
the calculations with different inputs. We show in Figure \ref{Fig04} the 
calculations with both Coulomb and medium corrections (solid curve),
calculations without any medium corrections (dashed lines), calculations that
exclude Coulomb distortions but keep medium corrections (dashed-dotted curve), and 
calculations without either Coulomb or medium corrections (dotted curve). 

The numerical results for the single particle cross sections with different configurations are shown 
in Table \ref{t_preAl23}. For each of the four configurations - $^{22}$Mg - 0$^{+}_{gs}$, 2$^{+}_{1}$, 
4$^{+}_{1}$, 4$^{+}_{2}$ - the corresponding relative differences between full calculations and 
calculations without Coulomb corrections are  found to be 15\%, 17\%, 19\% and 20\%, respectively, 
whereas between full calculations and calculations without medium corrections the corresponding 
percentage differences are found to be 24\%, 21\%, 16\% and 13\%, respectively.

\setlength{\tabcolsep}{0.1cm}
\begin{table*}[hbtp]
\begin{center}
\begin{tabular}{cccccccccc} 
\hline
    $J_\pi$       &  V$_0$ &  r$_0$  &   a$_0$ & V$_{s0}$  &r$_{s0}$ & a$_{s0}$ &    r$_c$      & $S_{eff}$  \\
                  &  (MeV) &  (fm)   &   (fm)  &  (MeV)    &   (fm)  &    (fm)  &    (fm)       &  (MeV) \\
\hline
$|^{10}$Be($J^\pi$)$\otimes\nu 2s_{1/2}\rangle$&&&&&&&&\\
\hline
0$^+_{(g.s)}$         & 61.13 & 1.21 & 0.52 & -  & -   & -  & 1.21  &    0.504 \\
\hline
$|^{14}$N($J^\pi$)$\otimes\pi nlj\rangle$&&&&&&&&\\
\hline
1$^+_{(g.s)}$(1p1/2)  & 48.36 & 1.19 & 0.60 & -  & -   & -  & 1.19  &    7.297 \\
1$^+_{(g.s)}$(1p3/2)  & 48.36 & 1.19 & 0.60 & -  & -   & -  & 1.19  &    7.297 \\
\hline
$|^{16}$B($J^\pi$)$\otimes\pi 1p_{3/2}\rangle$&&&&&&&&\\
\hline
  0$^+_{(g.s)}$ & 79.46  &    1.09  &   0.50  &   35.0  &  1.09 &  0.50  &  1.09   &    23.330 \\
   3$^-_1$      & 80.35  &    1.09  &   0.50  &   35.0  &  1.09 &  0.50  &  1.09   &    23.979 \\
   2$^-_1$      & 80.75  &    1.09  &   0.50  &   35.0  &  1.09 &  0.50  &  1.09   &    24.273 \\
   2$^-_2$      & 81.85  &    1.09  &   0.50  &   35.0  &  1.09 &  0.50  &  1.09   &    25.078 \\
   1$^-_1$      & 82.17  &    1.09  &   0.50  &   35.0  &  1.09 &  0.50  &  1.09   &    25.318 \\ 
   3$^-_2$      & 79.93  &    1.09  &   0.50  &   25.0  &  1.09 &  0.50  &  1.09   &    26.066 \\                           
\hline                                                                                            
$|^{22}$Mg($J^\pi$)$\otimes\pi1d_{5/2}\rangle$&&&&&&&&\\                                                  
\hline
0$^+_{(g.s)}$  & 54.60 &   1.18   &   0.60  &   5.0  &  1.18  & 0.60  &   1.18  &   0.141 \\
 2$^+$         & 56.96 &   1.18   &   0.60  &   5.0  &  1.18  & 0.60  &   1.18  &   1.388 \\
 4$^+_1$       & 60.67 &   1.18   &   0.60  &   5.0  &  1.18  & 0.60  &   1.18  &   3.449 \\
 4$^+_2$       & 64.07 &   1.18   &   0.60  &   5.0  &  1.18  & 0.60  &   1.18  &   5.434 \\
\hline
$|^{23}$O($J^\pi$)$\otimes\nu2s_{1/2}\rangle$&&&&&&&&\\
\hline
1/2$^+_{(g.s)}$& 42.40 &  1.27    &   0.70  &   -     &  -     & -    &1.27&    3.610 \\
\hline
$|^{32}$Mg($J^\pi$)$\otimes\nu nlj\rangle$&&&&&&&&\\
\hline
0$^+_{(g.s)}$($1d_{3/2}$)   &   -    &      -  &    -   &    -    &   -   &   -   &   -    &  2.21 \\
    3$^-$($2p_{3/2}$)       &  79.92 &    1.04 &   0.70 &   10.0  &  1.03 &  0.70 &  1.04  &  5.07 \\
    3$^-$($1f_{7/2}$)       &  86.63 &    1.04 &   0.70 &   10.0  &  1.03 &  0.70 &  1.04  &  5.07 \\
    2$^+_2$($2s_{1/2}$)     &  51.55 &    1.04 &   0.70 &   10.0  &  1.03 &  0.70 &  1.04  &  5.22 \\
\hline
\end{tabular}
\caption{Bound state potential parameters for the systems studied in the present work. $r_0$, $r_{s0}$ 
and $r_c$ are the reduced radius of the bound state potentials where $r_i = R_i/A_p^{1/3}$ ($i=0, s0, c$).
$S_{eff}$ is effective separation energy:
$S_{eff}=S_i+E_x^{core}$ where $i=proton$~or~$neutron$ and $E_x^{core}$ is core excitation energy.}
\label{bound}
\end{center}
\end{table*} 

\subsection{$^{9}$Be($^{15}$O,$^{14}$N)X at 56~MeV/u}
One-proton removal reaction from $^{15}$O on a Be target has been measured at 56~MeV/nucleon  and 
the total knockout cross section is reported as 80$\pm$20~mb in Ref. \cite{JEP04}. The authors were
able to explain the orbital occupancy of valence protons with a pure $1p_{1/2}$ 
single particle state using a Glauber reaction model. Their calculations imply that the $1p_{3/2}$ 
state could also have a small contribution because the calculations with only the 
$1p_{1/2}$ state yield a narrower momentum distribution than observed in the experiment. The
physical implication of this is a possible knockout from more deeply bound protons in the $1p_{3/2}$ 
state. The contributions from each of the $p$-states yield  spectroscopic factors of 1.27(9) and 0.100(75)
for the $1p_{1/2}$ and the $1p_{3/2}$ orbital, respectively (Ref. \cite{JEP04} and references therein).

\begin{figure}[ht]
\includegraphics[scale=0.35,keepaspectratio=true,clip=true]{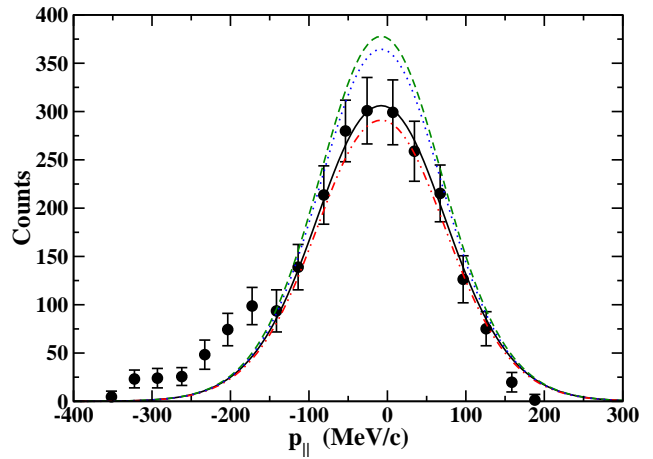}
\caption[Fig. 5]{(Color online) Longitudinal momentum distributions for the $^{9}$Be($^{15}$O,$^{14}$N)X 
reaction at 56~MeV/u. Solid lines represent calculations that include both Coulomb and medium corrections. 
Dashed lines stem from calculations that do not include medium corrections. Calculations denoted by 
dashed-dotted curves are performed without Coulomb corrections. The dotted curve does not include medium 
effects nor Coulomb corrections. The data are taken from  
Ref. \cite{JEP04}.} \label{Fig05}
\end{figure}

We have followed the interpretation of Ref. \cite{JEP04} and calculated the one-proton removal cross sections 
for the same reaction with the same orbital occupancy assumption. The parameters
are shown in Tables \ref{bound} and \ref{grdens}. Our calculations with both Coulomb and medium corrections 
by slightly changing the spectroscopic factors as 1.42 and 0.13 are in agreement with the results of Ref. \cite{JEP04}.
The calculated one-proton removal cross sections are 78.79~mb, 75.20~mb, 93.98~mb and 90.74~mb
with both Coulomb and medium corrections, no Coulomb corrections, no medium corrections and 
neither medium effects nor Coulomb corrections, respectively. The difference between full calculations, including medium and 
Coulomb scattering effects, and calculations without Coulomb corrections is of the order of 5\%, and between
full calculations and calculations without medium effects is nearly 19\%. 
This is remarkable even though it fits again within the error  of the total knockout cross section experimental data. 
We thus conclude that for this case, medium effects and Coulomb distortion do not have a sizable 
impact on the extraction of spectroscopic factors. However, one can easily see from Fig. \ref{Fig05} 
that the data shows an asymmetry which can only be explained with inclusion of higher-order effects 
in the reaction mechanism. Distortions will be manifest due to continuum-continuum coupling of states 
involving the interaction of core with the valence proton. These mechanisms have not been considered in the present work. 

\subsection{$^{12}$C($^{17}$C,$^{16}$B)X at 35~MeV/u}
\subsubsection{Transverse momentum distributions}
\begin{figure}[t]
\includegraphics[scale=0.35,keepaspectratio=true,clip=true]{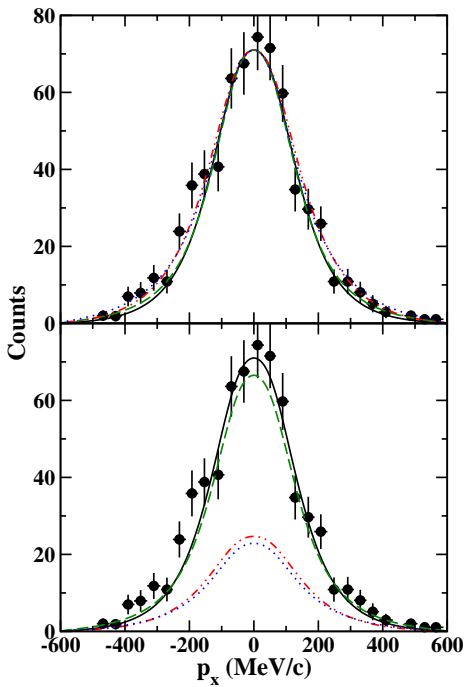}
\caption[Fig. 6]{(Color online). Transverse momentum distributions for the $^{12}$C($^{17}$C,$^{16}$B)X system
at 35~MeV/u. Solid lines represent calculations 
including both Coulomb and medium corrections. Dashed lines stem from calculations that do not include medium corrections. 
Calculations represented by dashed-dotted curves are performed without Coulomb corrections. The dotted curve does 
not include medium effects nor Coulomb corrections. The data are taken from Ref. \cite{LEC09}.
{\it Top panel:} One can see that, when properly scaled, all four curves from the calculations reproduce the shape of 
the momentum distributions.
{\it Bottom panel:} The relative differences of our results are illustrated when the full calculation (solid line)
is scaled to the data.}\label{Fig06}
\end{figure}

One-proton removal reaction from $^{17}$C, with a separation energy of 23~MeV, has been measured in 
the reaction $^{12}$C($^{17}$C,$^{16}$B)X at 35~MeV/nucleon with the goal to understand the low-lying structure 
of the unbound $^{16}$B nucleus. Using this reaction, Ref. \cite{LEC09} studied the unbound $^{15}$B+$n$ 
system with the assumption of a $d$-wave neutron decay. Our interest is to
compute the transverse momentum distribution of the $^{16}$B fragment
following the same assumptions  as in Ref. \cite{LEC09} in order to study the consequences of  
medium and Coulomb corrections. The configuration of the proton removed from $^{17}$C is assumed to be
\begin{eqnarray} 
&&|^{17}\text{C}\rangle = \alpha_1|^{16}\text{B}(0^-)\otimes \pi 1p_{3/2}\rangle \nonumber \\
&+& \alpha_2|^{16}\text{B}(3^-_1)\otimes \pi1p_{3/2}\rangle + \alpha_3|^{16}\text{B}(2^-_1)\otimes \pi1p_{3/2}\rangle \nonumber \\ 
&+& \alpha_4|^{16}\text{B}(2^-_2)\otimes \pi1p_{3/2}\rangle + \alpha_5|^{16}\text{B}(1^-_1)\otimes \pi1p_{3/2}\rangle \nonumber \\ 
&+& \alpha_6|^{16}\text{B}(3^-_2)\otimes \pi1p_{3/2}\rangle,
\end{eqnarray}
where $\alpha_i$ is the spectroscopic amplitude for a core-single particle configuration $i=(c\otimes nlj)$. 

Using spectroscopic factors obtained by means of a shell-model calculation with the WBP interaction 
\cite{WAR92}, Ref.  \cite{LEC09}  obtained a good agreement
between data and calculated transverse momentum distributions. But the measured total cross section
is 6.5(1.5)~mb against a theoretical result of 24.7~mb. The explanation of this large
difference is proposed in Ref. \cite{GAD04} as a reduction of the spectroscopic factor by 70\%  for
strongly bound nucleon systems. After this spectroscopic reduction is accounted for, the theoretical estimates 
for the cross section becomes 7.5~mb, in reasonable accordance with the data.

In the present work, we do not elaborate on the assumption introduced in Ref. \cite{LEC09}, and 
we use the same configuration and spectroscopic factors
as in \cite{LEC09}. The proton binding potential parameters are shown in Table~\ref{bound}
which are adjusted to obtain the effective separation energies. The ground state
densities are also listed in Table~\ref{grdens}. Here, as it is shown in Figure \ref{Fig06}, we find that
medium corrections change the total knockout cross sections by  5\%, but the Coulomb corrections 
have a very large effect which is almost 60\% between calculations with Coulomb and without Coulomb distortion. 
The reason for this difference is that the Coulomb distortion and repulsion effectively increases the collision 
distance at the small impact parameters needed to remove a strongly bound nucleon. This was not observed 
in the previous case ($^{9}$Be($^{15}$O,$^{14}$N)X at 56~MeV/u) because of the small nuclear binding in that case. 
We have also observed that this effect sharply reduces the calculated cross sections and the removal 
is more effective as the bombarding energy decreases. 

\subsubsection{Longitudinal momentum distributions}
We have made a more systematic analysis to understand  the reason of the effect discussed in the previous subsection.
We have observed that the strong dependence on Coulomb distortions are also present in longitudinal 
momentum  distributions. It has long been thought that longitudinal momentum distributions are free 
from uncertainties related to the knowledge of the optical nucleus-nucleus potentials when compared to the transverse 
distributions. This was first shown in Ref. \cite{BM92}.
Here we report calculations for  the same $|^{16}\text{B}(0^-)\otimes \pi 1p_{3/2}\rangle$ configuration, with the same
parameters and ground state densities, as discussed in the previous subsection. 
We find that although the Coulomb distortions create a similar 
effect for this particular knockout reaction on both transverse and 
longitudinal momentum distributions as can be seen in Fig.~\ref{Fig02}, 
the effect on transverse momentum distributions is bigger than the 
corresponding one for longitudinal momentum distributions by about 5\%. 
This is expected on physics grounds. Nonetheless, such a large effect on 
longitudinal momentum distributions was not initially anticipated. 
By comparison with other cases, we found that this large effect   
is due to the low bombarding energy in this particular reaction combined 
with a large binding energy of the projectile. This interpretation 
is also as it is validated by inspection of Figs.~\ref{Fig02}~and~\ref{Fig03}.

The source of this difference stems from the diffraction dissociation contribution to the cross sections.
To substantiate our claim, we have looked at the details of the knockout cross section which has two
parts for the production of a given final state of the residue. The most important of the two, commonly 
referred to as stripping or inelastic breakup, represents all events in which the removed nucleon reacts 
with and excites the target from its ground state. The second component, called diffractive or elastic 
breakup~\cite{BB88}, represents the dissociation of the nucleon from the residue through their two-body interactions 
with the target, each being elastically scattered. We notice that the total stripping cross 
section is given by \cite{BH04}:
\begin{eqnarray}
\sigma_{\mathrm{str}}  &  =&S(c;nlj) \frac{2\pi}{2l+1}\ \sum _{m}\int_{0}^{\infty}db_{n}%
\ b_{n}\ \left[  1-\left\vert S_{n}\left(  b_{n}\right)  \right\vert
^{2}\right]  \ \nonumber\\
&  \times&\int d^{3}r\ \left\vert S_{c}\left(
b_c\right)  \right\vert ^{2}\left\vert
\psi_{lm}\left(  \mathbf{r}\right)  \right\vert ^{2}, \label{strippxs}
\end{eqnarray}
whereas the integrated diffraction dissociation cross section is given by \cite{BG06}:
\begin{eqnarray}
\sigma_{\mathrm{dif}}    &=&S(c;nlj) \frac{2\pi}{2l+1}\sum _{m} \int_{0}^{\infty}db_{n}%
\ b_{n}  \nonumber\\
&  \times&\left\{ \int d^{3}r   \bigg\vert S_{n}\left(  b_{n}\right)S_{c}\left(
b_c\right) \psi_{lm}\left(  \mathbf{r}\right)\bigg\vert
^{2}\right.\nonumber\\
&-& \sum_{m'}\left. \left\vert \int d^3r  \psi_{lm'}\left(  \mathbf{r}\right)  S_{c}\left(
b_c\right)  S_n(b_n) 
\psi_{lm}\left(  \mathbf{r}\right)  \right\vert^{2} \right\}.
\label{sigdiff}
\end{eqnarray}

One can see from these expressions that the stripping cross sections are not affected by 
the Coulomb distortions because this distortion is manifest through a real phase in the eikonal 
S-matrices calculated in the Glauber approximation. The magnitude of the cross sections 
are therefore not changed, as the square of the S-matrices entering Eq. \eqref{strippxs} are only 
changed by the imaginary part of the potential entering Eq. \eqref{sib}. On the other hand, the second 
term of the diffraction dissociation cross sections in Eq. \eqref{sigdiff} is appreciably modified by 
the Coulomb phase factor. As seen from Figure \ref{Fig02}, the effect gets smaller with decreasing 
target atomic number because the Coulomb phase increases, or when the beam energy increases because 
then the Coulomb recoil becomes irrelevant. 

\subsection{$^9$Be($^{11}$Be,$^{10}$Be)X at 60~MeV/u}
\begin{figure}[t]
\includegraphics[scale=0.35,keepaspectratio=true,clip=true]{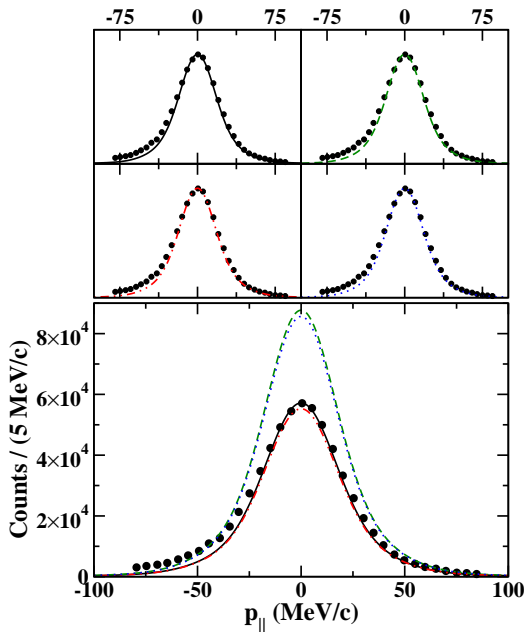}
\caption[Fig. 7]{(Color online). Longitudinal momentum distributions of for the reaction 
$^9$Be($^{11}$Be,$^{10}$Be) at 60~MeV/nucleon.  Solid lines represent calculations that 
include both Coulomb and medium corrections. Dashed lines stem from calculations that do not include
medium corrections. Calculations denoted by dashed-dotted curves are performed without Coulomb corrections. 
The dotted curve does not include medium effects nor Coulomb corrections. The data is taken from Ref.~\cite{AUM00}.
{\it Top panel:} One can see that, when properly scaled, all four curves from the calculations reproduce the shape of 
the momentum distributions.
{\it Bottom panel:} The relative differences of our results are illustrated when the full calculation (solid line)
is scaled to the data.}
\label{Fig07}
\end{figure}

In order to further understand the dependence of the Coulomb distortion on the nuclear binding, we 
consider the reaction $^9$Be($^{11}$Be,$^{10}$Be)X at 60~MeV/u which can be modeled by a core plus 
valence system with the assumption $^{10}$Be$_{gs}(0^+)+{\rm n}$ in $2s_{1/2}$ orbital for the ground 
state of $^{11}$Be$_{gs}(1/2^+)$ ($S_n=0.504$~MeV). Here we use the same Woods-Saxon potential parameters 
for the bound state as published in Ref. \cite{hen96}: ($R_0=2.70$ fm, $a_0=0.52$ fm). In Figure~\ref{Fig07} 
and Table \ref{cross} we present our results for the neutron removal longitudinal momentum distribution 
of 60~MeV/nucleon $^{11}$Be projectiles incident on $^9$Be targets. 

\begin{table}[hbtp]
\begin{center}
\setlength{\tabcolsep}{0.06cm}
\begin{tabular}{|c|ccc|ccc|} 
\hline
$\sigma_{-1n}$&\multicolumn{3}{c|}{$^{12}$C($^{17}$C,$^{16}$B)}&\multicolumn{3}{c|}{$^9$Be($^{11}$Be,$^{10}$Be)}\\
$=\sigma_{dif}+\sigma_{str}$ &  Full    & no Coul   & no med  &   Full    & no Coul    & no med \\
\hline
Strip. [mb]               &  7.10 &  7.10 &\hspace{5px}5.09 & 126.5 &  126.5 &  169.7  \\ 
Diff.  [mb]               & 18.63 &  2.42 & 19.39 &\hspace{5px}52.8 &\hspace{5px}46.7 &  104.8  \\ 
Total  [mb]               & 25.74 &  9.52 & 24.48 & 179.3 &  173.2 &  274.5  \\
\hline
\end{tabular}
\caption{The cross sections calculated for the systems, $^{12}$C($^{17}$C,$^{16}$B) at 35~MeV/nucleon and 
$^9$Be($^{11}$Be,$^{10}$Be) at 60~MeV/nucleon.}
\label{cross}
\end{center}
\end{table} 

It is evident from Fig. \ref{Fig01} that $^{17}$C has the smallest ``effective" size and that $^{11}$Be
has the biggest size among the low energy systems in this study.  The nuclear size 
is important for low energy cases because the diffraction dissociation becomes dominant when the nuclear 
size is smaller, but the stripping becomes dominant when the nuclear size is bigger.
The reason for this is that a large projectile feels the nuclear interaction already 
at large impact parameters. A small projectile can come closer to the target where the Coulomb interaction
is larger. The evidence of this can be seen in Table \ref{cross}. It is thus clear why medium and 
Coulomb corrections are more important in the $^9$Be($^{11}$Be,$^{10}$Be) and the $^{12}$C($^{17}$C,$^{16}$B) 
cases, respectively.

\subsection{$^{12}$C($^{24}$O,$^{23}$O)X at 920~MeV/u}
The momentum distribution of the one-neutron removal residues from the $^{12}$C($^{24}$O,$^{23}$O)X reaction was
measured for the first time at 920~MeV/nucleon and reported in Ref. \cite{KAN09}. The data could be explained 
with a spectroscopic factor $S$=1.74(19) of an almost pure 2$s_{1/2}$ 
{single-particle} state for the valence neutron. This work,
together with recent theoretical calculations, suggests that $^{24}$O is a newly discovered doubly magic nucleus.
The one-neutron removal cross section was found to be 63(7) mb. The
calculations in Ref. \cite{KAN09} were based on a few-body Glauber formalism \cite{OGA94} for two 
configurations: (a) $^{23}$O$_{gs}(1/2^+)+{\rm n}$ in $2s_{1/2}$ orbital and (b) $^{23}$O$_{gs}(5/2^+)+{\rm n}$ in $1d_{5/2}$ orbital. 
The wave functions for the configurations are obtained with a Woods-Saxon potential by
adjusting the depth of the potential to reproduce the one-neutron separation energy 
$S_n$=3.61(27)~MeV \cite{AUD03}. Using a pure 2$s_{1/2}$ configuration with $S$=1 leads to
a cross section of 34~mb. The calculation is in agreement with the data
when it is multiplied by $S$=1.74(19). This large spectroscopic factor indicates
that the single-particle strength of the valence neutron is strongly weighted in the 2$s_{1/2}$ state.

In the present work we have reproduced the data of Ref. \cite{KAN09} also by assuming a 2$s_{1/2}$ orbital only.
The potential parameters for the bound state wave function are given in Table \ref{bound} and the ground state density for the
$^{23}$O$_{gs}$ core is obtained using liquid droplet model (LDM) densities \cite{MS74}, as indicated 
in Table \ref{grdens}.  To understand the differences between  medium effect models, four different
calculations including Coulomb corrections have been made for this system. The calculated  
one-neutron removal cross sections are 58.58~mb, 54.08~mb, 78.74~mb and 53.25~mb using 
free \cite{BC10}, Pauli corrected (Eq. 2), Brueckner (Eq. 3) and phenomenological parameterizations 
\cite{Xian98} of the nucleon-nucleon cross sections, respectively. Except for the result obtained with the  Brueckner 
theory, they are all in agreement with the previous work and with the data. 
The relative difference between the results obtained using Brueckner corrections and with free 
nucleon-nucleon cross sections is about 34\%. However, we do not consider a real discrepancy, as 
the Brueckner parametrization have been extrapolated well beyond their validity. 
Brueckner calculations are limited by the pion-production threshold, and should 
only be valid for projectile energies below 300~MeV/nucleon.  

Thus we verify that the experimental data for the reaction $^{12}$C($^{24}$O,$^{23}$O)X at 
920~MeV/u is well reproduced with the use of free nucleon-nucleon cross sections. The changes 
introduced by Pauli-blocking  with the geometric model are small, and the phenomenological account 
of medium effects at this high energy also basically agree with the results using free cross sections. 
\subsection{$^{12}$C($^{33}$Mg,$^{32}$Mg)X at 898~MeV/u}
\begin{figure}[ht]
\includegraphics[scale=0.35,keepaspectratio=true,clip=true]{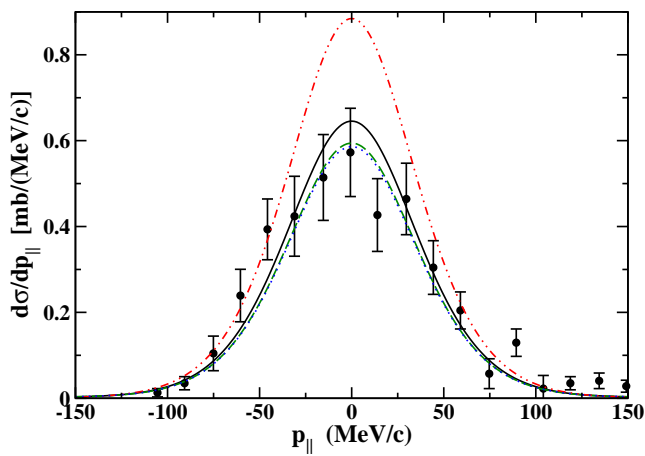}
\caption[Fig. 8]{(Color online). The longitudinal momentum distributions for 
$^{12}$C($^{24}$O,$^{23}$O)X reaction at 920~MeV/nucleon. The curves are calculated with the free 
NN cross sections (solid), with a geometrical account of Pauli blocking (dashed), a phenomenological 
fit from Ref. \cite{Xian98} (dotted), and a correction from Brueckner theory (dashed-dotted). 
The data has been taken from Ref.~\cite{KAN09}.}
\label{Fig08}
\end{figure}

The ground state structure of $^{33}$Mg, a nucleus belonging to the $N=20$ island of inversion, has 
been studied  in Ref.  \cite{KAN10} by means of nucleon-removal reactions on a carbon
target at 898~MeV/nucleon. The longitudinal momentum distribution of the $^{32}$Mg core was measured 
and the one-neutron removal cross section was found to be 74(4)~mb. Most of the
contribution to the ground state structure of $^{33}$Mg was shown to arise from the
$2p_{3/2}$ orbital.

The longitudinal momentum distribution obtained in Ref. \cite{KAN10}
cannot be reproduced with a pure single particle state. It has been discussed in details
in Ref. \cite{KAN10} the reason why a configuration mixing of different
single-particle states is needed. Two different configuration mixings for the ground state
of $^{33}$Mg were assumed. The first one is
\begin{eqnarray} 
&&|^{33}\text{Mg}_{gs}(3/2^-)\rangle = \nonumber \\
&& \alpha_1|^{32}\text{Mg}(2^+_1)\otimes \nu2p_{3/2}\rangle + \alpha_2|^{32}\text{Mg}(1^-)\otimes \nu2s_{1/2}\rangle  \nonumber \\
&+& \alpha_3|^{32}\text{Mg}(2^+_1)\otimes \nu1f_{7/2}\rangle  + \alpha_4|^{32}\text{Mg}(1^-)\otimes \nu1d_{3/2}\rangle  \nonumber \\
&+& \alpha_5|^{32}\text{Mg}(gs)\otimes \nu2p_{3/2}\rangle \label{fconf}
\end{eqnarray}
and the second is
\begin{eqnarray} 
&&|^{33}\text{Mg}_{gs}(3/2^+)\rangle = \alpha_1|^{32}\text{Mg}(3^+)\otimes \nu2p_{3/2}\rangle \nonumber \\
&+&\alpha_2|^{32}\text{Mg}(2^+_2)\otimes \nu2s_{1/2}\rangle +\alpha_3|^{32}\text{Mg}(3^+)\otimes \nu1f_{7/2}\rangle \nonumber \\
&+&\alpha_4|^{32}\text{Mg}(gs)\otimes \nu1d_{3/2}\rangle, \label{sconf}
\end{eqnarray}
where $\alpha_i$ are the spectroscopic amplitudes for each single-particle orbital.
The values of the corresponding spectroscopic factor $S_i$ were found by $\chi^2$ minimization and their values 
for the second configuration are $S_1=2.2^{+0.2}_{-0.5}$, $S_2=0.1^{+0.0}_{-0.1}$, 
$S_3=1.1^{+0.1}_{-0.5}$ and $S_4=0.0^{+0.5}_{-0.0}$ \cite{KAN10}.

\begin{figure}[ht]
\includegraphics[scale=0.35,keepaspectratio=true,clip=true]{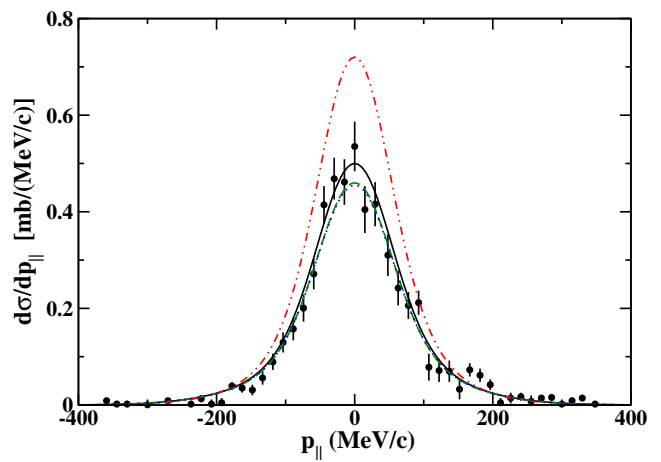}
\caption[Fig. 9]{(Color online). The inclusive longitudinal momentum distributions for 
the $^{12}$C($^{33}$Mg,$^{32}$Mg)X system at 898~MeV/nucleon. 
The data has been taken from R. Kanungo {\it et~al.} \cite{KAN10}. The curves are calculated with the 
free NN cross sections (solid), with a geometrical account of Pauli blocking (dashed), a 
phenomenological fit from Ref. \cite{Xian98} (dotted), and a correction from Brueckner theory (dashed-dotted).}
\label{Fig09}
\end{figure}

In our calculations we have chosen the second configuration set used in Ref. \cite{KAN10}, Eq. \ref{sconf}, since
the $^{33}$Mg ground state is usually accepted to be $J^\pi=3/2^+$.
We apply the same procedure as described before
to obtain bound state wave functions and eikonal phases. The parameters
for the bound state potentials and ground state densities are 
shown in Table \ref{bound} and Table \ref{grdens}, respectively.
We have used nearly the same spectroscopic factors within the error bar range
of Ref. \cite{KAN10} to make a consistent comparison of the 
medium effects. Our results yield a small but relevant variation
of the one-neutron removal cross sections using the free, Pauli corrected, and
phenomenological NN cross sections, namely 83.70~mb, 77.90~mb, and 77.63~mb, respectively. 
As observed in the case of the $^{12}$C($^{24}$O,$^{23}$O)X at 920~MeV/u, the use of Brueckner 
corrected NN cross sections yields 112.92~mb, about 35\%
relative to the calculations using the free NN cross sections. For the same reason as with
the previously considered reaction, this discrepancy is meaningless as one extrapolates the 
Brueckner results beyond their regime of validity. 

Both reactions considered above are very illustrative as they show a great consistency between 
the calculations performed by different authors, with somewhat different methods. They also show the 
expected relevance of medium corrections at intermediate and low energy collisions.
\subsection{Relevance of medium effects}
As we mentioned above, medium effects have been routinely neglected in the experimental analysis of
knockout reactions. But their relevance has been known for a long time in the analysis of elastic and inelastic scattering, as well as
of total reaction cross sections \cite{Ra79,HUS91}. The effects are larger at lower bombarding energies, where Pauli blocking strongly
reduces the nucleon-nucleon cross sections in the medium. A systematic study of these effects has been presented in Ref. \cite{HUS91}.

\begin{figure}[t]
\includegraphics[scale=0.33,keepaspectratio=true,clip=true]{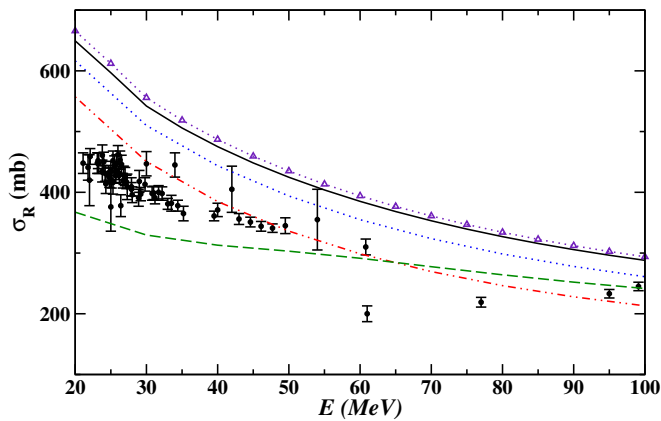}
\caption[Fig. 10]{(Color online). The total reaction cross section of the p\;+$\;^{12}$C taken from 
Ref.~\cite{CAR96}. The curves are calculated with the free NN cross sections from Ref.~\cite{JTWT93} 
(solid), with a geometrical account of Pauli blocking (dashed), a phenomenological fit from 
Ref.~\cite{Xian98} (dotted), and a correction from Brueckner theory (dashed-dotted).
The triangle-dotted curve is calculated with the same free NN cross sections from Ref.~\cite{JTWT93}, 
but with an another HFB calculation \cite{Bei74} for the $^{12}$C ground state density.}
\label{Fig10}
\end{figure}

To corroborate these statements, in figure \ref{Fig10} we show the data on p + $^{12}$C reaction cross sections 
taken from the Ref.~\cite{CAR96} in the energy region of our interest, 20-100 MeV/nucleon.   The cross sections were 
calculated from the relation
\begin{equation}
\sigma_R=2\pi \int db \ b \left[ 1- \left| S(b)\right|^2 \right],
\end{equation}
where $S(b)$ has been calculated using Eqs. (\ref{sib},\ref{eikphase},\ref{fnn}) and the carbon 
matter density from a Hartree-Fock-Bogoliubov calculation \cite{Bei74}.  Several distinct calculations are shown. 
The solid curve uses Eq. \eqref{eikphase} with the free nucleon-nucleon cross sections and the carbon matter 
density from a HFB calculation \cite{Nus12}, whereas the triangle-dotted  curve (the triangles are not data, 
but used for better visibility) uses a different HFB density \cite{Bei74},  consistent with the calculations 
presented in Ref. \cite{Tra02}. As expected, that the agreement between the two calculations is very good. 

The other curves in figure \ref{Fig10} show the same calculation procedure, but  including medium 
corrections for the nucleon-nucleon cross section. The results are evidently very different than the previous 
ones. The dotted (dashed-dotted) [dashed] curves use phenomenological (Brueckner) [Pauli geometrical] recipes 
for medium effects on the cross sections. Based on the large error bars and spread of the experimental data, 
it is hard to judge what model adopted for medium corrections yields the best agreement with the data. 
It is clear that the inclusion of medium effects change the results to yield a closer reproduction of the data.  

These findings are in agreement with our present understanding of medium modifications of the reaction 
cross sections and of several other reaction channels involving heavy ion scattering at intermediate 
energies ($\sim 50$ MeV/nucleon).

\bigskip
\section{Conclusions}
Often neglected effects, such as medium modifications of the nucleon-nucleon cross 
sections and Coulomb distortion, modify appreciably the nucleon knockout cross sections.
As we have shown, these effects do not lead to an appreciable modification of the shapes of momentum distributions.  
This is explained by the fact that the momentum distributions are largely the Fourier transforms of the contributing 
parts of the single-particle wavefunctions, overwhelmingly their asymptotic regions, which are the Whittaker functions for 
protons or the Hankel functions for neutrons, sensitive only to the orbital momentum and the nucleon binding energies.
We have shown these features explicitly by comparing our results with a large number of available experimental data. 
As expected on physics grounds, these corrections are  larger for experiments at  
lower energies, around 50~MeV/nucleon, and for heavy targets. 

As more experiments make use of heavier targets, it is worthwhile to illustrate the relevance of 
Coulomb corrections. Medium effects in knockout reactions have also been frequently ignored in the past. We show that they 
also have to be included in order to obtain a better accuracy of the extracted spectroscopic factors. 
Although these conclusions might not come as a big surprise, they have not been properly included in 
many previous experimental analyses.  

\section*{Acknowledgements}
This work was partially supported by the US-DOE grants DE-SC004972 and DE-FG02-08ER41533 and 
DE-FG02-10ER41706, and by the Research Corporation. M. Karako\c{c} has also been partially 
supported by the TUBITAK grant 109T373.

\end{document}